# Resolving the Binding Constraint on Circular Economy: Principal Return Rate as Interest-Free Monetary Architecture

*Amir Rashid*
*KTH Royal Institute of Technology, Stockholm- Sweden*
*Email:* amirr@kth.se
ORCID: 0000-0002-5960-2159

## Abstract

Green growth through circular economy is the dominant institutional response to ecological breakdown. The evidence against its sufficiency is structural across three dimensions: material flow arithmetic prevents loop closing at scale in a growing economy; thermodynamic constraints make completely closed material loops physically impossible; and circular material flows account for only 1.4% of global GDP while CE-principled service activities already generate approximately two thirds of GDP without arresting ecological overshoot. The constraint therefore is not at the level of production organisation but at the level of the monetary architecture that structurally compels resource throughput expansion regardless of how efficiently or sustainably that production is organised. Six independent scholarly traditions including ecological economics, post-Keynesian monetary theory, institutionalist political economy, heterodox monetary reform, the Gesellian demurrage tradition, and the historical critique of usury have converged on the interest as the central structural problem in any economy that does not require perpetual growth. Yet no existing proposal has specified a technically coherent mechanism to replace the monetary control function that interest currently performs, a gap unfilled across nine decades of sovereign money proposals. This article introduces the Principal Return Rate (PRR) as that mechanism. The PRR replaces interest not as a cost of credit but as a rate of return instrument: sovereign money, anchored to productive demand as in the current system, is returned to central bank reserves at rates determined by the PRR. Inflation is controlled through central bank PRR adjustment, directly analogous to interest rate policy but acting on return velocity rather than borrowing cost, and without the cost-push inflationary channel that interest itself generates. The article makes one original theoretical contribution: the PRR system closes the functional design gap that has remained open across nine decades of sovereign money proposals and six independent scholarly traditions. It additionally introduces the hypothesis that positive time preference is partially endogenous to the interest-bearing system, generated through the inflation that interest produces, making its conventional justification partially circular.

*Keywords*: Principal Return Rate, monetary growth imperative, sovereign money, circular economy, interest-free monetary system, productive capital, Banking-as-a-Service, polycentric governance, ecological macroeconomics, inflation control, time preference

# 1 Introduction

## 1.1 The green growth contradiction

As ecological breakdown intensifies, from accelerating biodiversity loss and resource depletion to the mounting costs of climate disruption, the economic system has come under increasing scrutiny as both its primary cause and the arena in which solutions must be found. This has framed an increasingly important and urgent question: can an economy that must grow to function also be an economy that respects planetary limits? The answer that has achieved broadest institutional consensus, embedded in the European Commission's Circular Economy Action Plans, the OECD's sustainability frameworks, and corporate transition strategies across sectors, is green growth, for which the circular economy has emerged as the dominant policy framework. It is worth noting that this is not a fringe proposition. It commands policy budgets measured in hundreds of billions of euros and provides the operational content of most national sustainable development commitments. Yet the empirical case for decoupling on which this framework rests has been substantially challenged (Parrique et al. 2019).

This challenge is part of a broader scholarly assessment of the mainstream answer that is increasingly unfavourable, and the critique is structural rather than incidental, operating across three distinct analytical dimensions. First, at the level of material flows, Krausmann et al. (2017) demonstrate that global socioeconomic material stocks have risen 23-fold over the 20th century and currently require approximately half of all annual resource extraction simply for maintenance and expansion. With only 12% of stock inflows coming from recycling, the arithmetic of material flows structurally prevents CE from closing loops at meaningful scale within a growth-oriented economy. Second, at the level of engineering and thermodynamics, Korhonen et al. (2018), identify that the laws of thermodynamics make completely closed material loops physically impossible meaning virgin raw materials will always be required even to maintain the status quo. Third, and most significantly, at the macroeconomic level, the limitation of CE extends well beyond its material flow fraction. A holistic view must account for both fractions of the circular economy. The stock fraction already accounts for approximately two thirds of GDP in advanced economies, while the secondary material flow fraction accounts for only 1.4% of global GDP (Rashid 2026). The economy has implicitly conducted the circular economy experiment at dominant scale across both dimensions, and the ecological outcomes fall far short of what green growth promises. To date, CE has fallen short of its promise to reduce resource demand (Suski et al. 2023), framed as promising the radical but delivering the familiar pathway of business-led economic growth (Clube & Tennant 2020).

The scholarly response has been to argue that CE must be integrated with post-growth or degrowth frameworks. Bauwens (2021) argues that CE is structurally incompatible with economic systems designed to pursue endless growth. Zink and Geyer (2017) identify the specific market mechanism by which this incompatibility manifests: circular economy activities operating in profit-maximising markets systematically produce rebound rather than primary production displacement, since the profit-maximising strategy is to grow overall production rather than substitute for primary. These contributions correctly identify that CE's limitations cannot be resolved through better production design alone. However, they share a critical gap since they identify the incompatibility between CE and growth-oriented systems without specifying the mechanism through which the growth imperative operates independently of the production model.

Among the multiple structural imperatives driving economic expansion, the interest-bearing architecture of money creation occupies a unique position. It operates above and independently of the production model, requiring aggregate growth regardless of what entrepreneurs choose to produce, what consumers choose to consume, or how well circular economy designers close material loops. The service sector paradox makes this conclusion unavoidable: if CE-principled activities already dominate economic value generation and yet cannot arrest ecological overshoot, then the binding constraint must be operating at a level that production organisation, however circular, cannot reach.

## 1.2 The monetary root cause

The growth imperative operating on the circular economy cannot lie in the production model. It must lie in the monetary system within which all production models operate. The specific mechanism is the interest-bearing structure of money creation. When every unit of money entering the economy is created as interest-bearing debt, the monetary architecture requires the economy to generate more money than was ever created simply to service the debt through which all money comes into existence. This compulsion operates structurally and independently of the production model. It persists regardless of how circular, how efficient, or how sustainably organised that production model becomes.

The evidence for this mechanism extends well beyond the CE domain. Fuders and Guerrero (2026), demonstrate through natural resource management cases that the monetary growth imperative applies to public and natural goods alike. This systematically favours liquidation over stewardship and financial return over ecological regeneration, regardless of the production model in place. Their natural resource cases and the CE service sector paradox are manifestations of the same structural dynamic at different scales. In both cases, the monetary architecture overrides the logic of long-term stewardship that ecological sustainability requires. This convergence is not confined to empirical observation only. Scholarly traditions including ecological economics, post-Keynesian monetary theory, institutionalist political economy, heterodox monetary reform, the Gesellian demurrage tradition, and the historical critique of usury have each made interest the central variable in their account of whether and how growth becomes structurally necessary. Yet no existing proposal has specified a technically coherent mechanism to replace the monetary control function that interest currently performs. This is the gap the present article fills.

## 1.3 The central claim and article structure

This article makes one theoretically original contribution: the Principal Return Rate (PRR) as the first technically coherent replacement for the monetary control function that interest currently performs. The PRR as a rate instrument, combined with the institutional architecture within which it operates, constitutes the PRR system. The PRR system thus closes a functional design gap that has remained open across nine decades of sovereign money proposals, from the Chicago Plan of the 1930s through to the most recent systematic reviews of the field (Keysser et al. 2025; Fuders & Guerrero 2026). The gap persists across the traditions mapped in Section 2 that have converged on interest as the central problem without converging on its solution. The article's premise that this gap urgently needs filling is evidenced by Rashid (2026). This empirical evidence shows that CE-principled activities already dominate economic value generation while ecological overshoot persists, establishing the monetary architecture, rather than the production model, as the binding constraint that CE operating within an interest-bearing system cannot overcome.

The article proceeds as follows. Section 2 maps the convergence of scholarly traditions on the interest rate as the central problem. Section 3 decomposes the bundled functions of interest to establish that the bundling is contingent rather than necessary. Section 4 introduces and specifies the PRR mechanism. Section 5 presents the institutional architecture within which the PRR operates. Section 6 addresses known and anticipated critiques. Section 7 concludes with acknowledged limitations and a four-stream future research agenda.

The article employs a theoretical framework methodology combining systematic literature synthesis across the six scholarly traditions, analytical decomposition of the interest rate's bundled functions, and constructive institutional design grounded in the decomposition's findings. The PRR mechanism is constructed from the logical requirements established by the decomposition rather than derived by modifying existing proposals, enabling a genuine departure from the functional architecture of interest rather than a modification of it.

### 1.4 Empirical and intellectual urgency

The urgency of monetary system reform through the PRR is reinforced by the empirical record of green growth policy failure. Decades of commitment to green growth ambitions have produced no demonstrable absolute decoupling of economic output from ecological impact, and ecological overshoot continues to accelerate despite CE-principled activities dominating economic value generation. The monetary constraint identified in the preceding sections has operated unimpeded throughout, not because green growth policy was insufficiently ambitious, but because it has operated entirely within the monetary architecture without challenging the structural compulsion to expand.

The practical window for this reform has never been wider. 146 countries and currency unions representing over 98% of global GDP are building sovereign digital money infrastructure (Atlantic Council 2025; BIS Papers No. 159, 2025). This development is driven not by a principled theoretical case for sovereign money creation but by a tactical defensive response to the rise of private digital currencies. The PRR is proposed at the precise moment when the architectural decisions that will determine whether this infrastructure reproduces or resolves the monetary growth imperative, are yet to be made.

Empirical evidence from the most significant economic disruptions of the past four decades further confirms both the necessity and the feasibility of this architectural shift. The 1991 Swedish banking crisis and the 2008 global financial crisis were both triggered by the collapse of credit booms fuelled by interest-bearing lending. In the case of Sweden, this occurred through deregulated bank competition that prioritised credit expansion over prudent lending, followed by a sharp rise in real interest rates (Englund 1999); in 2008, through securitised mortgage debt (Danielsson, Macrae, Vayanos & Zigrand 2020). Both confirm that systemic financial catastrophe is a recurring feature of interest-dependent monetary systems rather than an exceptional event. The COVID period reinforced this diagnosis from two directions simultaneously. First, the involuntary halt to economic activity in 2020 produced one of the largest single-year CO2 emissions drops on record, without causing systemic financial collapse (Le Quéré et al. 2021). Second, the recovery that followed relied on conventional monetary and fiscal stimulus rather than any CE-aligned or green-growth-targeted intervention, and emissions rebounded to within 1% of 2019 levels by 2021 (Davis et al. 2022). This dynamic illustrates the deeper argument advanced here. The service sector paradox showed that growth-throughput coupling persists even where circular economy principles are already dominant. The COVID episode shows the same coupling at the level of the aggregate economy, with no reference to circular economy or production model at all. Together, the two cases suggest that the coupling originates in the monetary architecture itself, rather than in how any particular production model, circular or otherwise, happens to be organised.

The scholarly record mirrors the empirical one: the diagnosis is established but the prescription has remained absent. Ninety years of sovereign money proposals since the Chicago Plan of the 1930s have failed to specify a functional replacement for interest as the monetary control mechanism. Keysser et al. (2025) identify 112 growth imperative mechanisms across 248 publications without identifying a monetary control replacement. Fuders and Guerrero (2026) discard five categories of counterarguments while leaving the replacement mechanism question unresolved confirming that the scholarly diagnosis has deepened considerably while the prescription remains completely absent.

## 2 The growth imperative debate: convergence without resolution

### 2.1 The strong claim: interest as structural growth imperative

The most direct formulation of the monetary growth imperative comes from Binswanger (2009, 2015), who established the foundational argument that positive growth rates are structurally necessary in a capitalist economy with private bank money creation. Commercial banks must retain earnings to maintain capital adequacy buffers, creating a permanent drain on the circular flow of money that can only be replenished through new credit expansion and hence through economic growth. The growth imperative is not a preference of individual actors but a structural consequence of the institutional arrangement through which money enters the economy.

Farley et al. (2013) extend this argument into the ecological domain, arguing that current monetary institutions leave society facing an unacceptable choice between ecological catastrophe and human misery. They propose that this choice is escapable only through a transition to a steady-state economy with debt free sovereign money creation. The ecological economics tradition more broadly identifies that interest rates compel the liquidation of natural capital whenever financial returns on extracted resources exceed the growth rate of the living system. This dynamic that Fuders and Guerrero (2026) demonstrate applies to public and natural goods across multiple empirical domains. Their systematic examination and rejection of five categories of objections to the growth imperative thesis represents the most recent and comprehensive defence of the strong claim in the current literature.

## 2.2 The moderate claim: interest as institutional growth driver

Svartzman et al. (2020) situate the problem institutionally rather than mechanistically. The progressive generalisation of interest-bearing debt money since the thirteenth century has produced an economic architecture in which growth is not merely incentivised but structurally expected, rewarded, and when absent considered destabilising. Whether the growth imperative is hardwired into monetary mechanics or deeply encoded in the institutional arrangements surrounding those mechanics, the consequence for ecological sustainability is the same. Svartzman et al. conclude that monetary contestations, challenges to the structural role of interest, are critical to achieving any post-growth socio-economic system. This is the institutionalist tradition's own call for architectural reform of money's role in growth, made independently of any specific replacement mechanism.

## 2.3 The optimistic rebuttal: stationarity is conditionally possible with positive interest

The most sustained scholarly challenge to the growth imperative thesis comes from post-Keynesian stock-flow consistent modelling that demonstrates, under specific conditions, the compatibility of positive interest rates with a stationary economy[1]. Jackson and Victor (2015) show that a stationary state, defined here as zero net growth in financial wealth and productive capital stocks rather than zero consumption, is reachable without eliminating interest. This requires that households spend down interest income and existing wealth rather than reinvesting it, a condition the authors term consumption out of wealth, combined with countercyclical fiscal intervention to compensate for interest-driven demand shortfalls. Richters and Siemoneit (2017), examining five post-Keynesian models, find that a stationary state with zero net additions to stocks of financial wealth and productive capital is achievable with positive interest rates if consumption out of wealth exceeds a threshold rising with the interest rate. Cahen-Fourot and Lavoie (2016) confirm compatibility under certain parameter configurations, and Hein and Jimenez (2022) demonstrate theoretical achievability but only under specific redistributive conditions that do not emerge spontaneously from market dynamics.

---

[1] This article uses the term 'stationary economy' to mean an economy freed from the monetary compulsion to expand resource throughput and ecological impact. The concern is specifically with the monetary compulsion that makes resource throughput expansion and ecological impact structurally inevitable regardless of production model choices, not with the growth of productive capital actively deployed in socio-ecologically purposeful activity.

This body of work reframes rather than dismisses the growth imperative problem. But the conditions required such as deliberate redistribution of profits, managed consumption out of wealth, sustained countercyclical fiscal intervention, are themselves deeply problematic. They face four compounding limitations: they are behaviourally unrealistic as sustained policy targets; they generate inflationary aggregate demand pressure in an economy whose output is not growing; they are empirically undemonstrated within existing positive interest economies despite decades of fiscal activism; and they reproduce the consumerism logic that ecological economics seeks to transcend. More fundamentally, fiscal compensation financed through borrowing adds new interest-bearing debt that regenerates the monetary compulsion it seeks to offset, making the stationary state structurally self-defeating within an interest-bearing architecture.

The empirical record provides the most decisive challenge to the optimistic position. Hartley and Kallis (2021), examining ten historical cases across five millennia and multiple civilisations, find that no society ever imposed the zero net saving condition that the theoretical models require. In seven of the ten examined cases, interest-bearing loans in slow or non-growing economies produced unpayable debts, debtor dispossession, and social upheaval. The condition that makes positive interest theoretically compatible with stationarity has never been implemented in any economy across recorded human history. The optimistic camp has demonstrated that stationarity is theoretically possible with positive interest. However, it has not shown that such approach to stationarity is feasible since the conditions necessary for it are not only demanding, but they also appear unrealistic, inflationary, empirically ineffective, and ecologically contradictory all at once. The question this article takes up is deeper and challenging. What monetary architecture makes stationarity a default behaviour of the system, rather than an achievement that depends on infeasible or fragile conditions to hold indefinitely?

### 2.4 Heterodox monetary reform: the sovereign money tradition

A fourth tradition approaches the growth imperative through the specific institutional critique of private bank money creation. The Chicago Plan, advanced by University of Chicago economists in 1933 and most prominently articulated by Irving Fisher (1936), envisaged 100% reserve backing for deposits. It correctly identified private bank money creation as the structural source of financial instability and proposed its transfer to sovereign control but treated interest itself as unproblematic and retained it as the monetary control instrument. The most rigorous modern revival by Benes and Kumhof (2012) and its extension in Kumhof (2025) at the IMF finds support for all of Fisher's original claims. However, it resolves the control question by administering multiple interest rates rather than by replacing interest as the control instrument. The contemporary sovereign money movement, represented most prominently by Positive Money (Jackson & Dyson 2012), ultimately falls back on market driven interest rate adjustment as the residual monetary control mechanism. The heterodox monetary reform tradition has correctly identified the problem and proposed sovereign money as its solution in principle but has not specified the functional replacement for interest as the monetary control instrument. This gap remains open across nine decades of sovereign money proposals.

## 2.5 The historical critique of usury: millennial convergence

The fifth tradition is the oldest and broadest. Across radically different civilisations, the prohibition of usury appears independently in Hebrew law, Platonic philosophy, Roman legislation, Indian Dharmashastra, Chinese imperial edicts, Christian canon law, and Islamic jurisprudence. The significance of this convergence is epistemic rather than normative. That lawmakers, philosophers, theologians and jurists in cultures separated by geography, language, and epoch independently identified the same structural problem constitutes historical evidence of a kind that neither theoretical modelling nor single-society empirical work can supply. The Hartley and Kallis (2021) findings discussed above are one component of this record. What the wider tradition reveals is that interest bearing money consistently generates extractive dynamics which reflective observers across millennia recognised as structurally problematic well before modern monetary economics existed.

## 2.6 The Gesellian tradition: a sixth stream

A sixth tradition approaches the growth imperative through the lens of money's function as a store of value. Silvio Gesell's original insight was that interest-bearing money systematically privileges liquidity over productive investment and short-term financial return over long-term stewardship because money held earns interest while real assets depreciate. Fuders (2026) develops this argument into a systematic dual microeconomic and macroeconomic critique, proposing Gesellian demurrage as the corrective mechanism. Demurrage in this context means a periodic holding charge on money designed to discourage hoarding by making idle cash holdings costly. The Gesellian tradition provides the most direct microeconomic grounding for the growth imperative argument. It shows that the compulsion operates at the level of individual investment decisions in which interest bearing financial assets systematically outcompete long-horizon productive stewardship. However, Gesellian demurrage confirms the diagnostic without providing the prescription. It addresses the hoarding incentive and short-termism bias but does not specify a functional replacement for interest as the monetary supply control instrument.

## 2.7 What the debate reveals: the convergence point

Across six independent scholarly traditions, the growth imperative debate has generated a remarkable degree of convergent diagnosis. No major school argues that interest is irrelevant to the design of a stationary economy. The disagreement is about the strength and conditionality of the relationship, not its existence. Keysser et al. (2025), reviewing 248 publications and identifying 112 growth imperative mechanisms across 21 thematic clusters, confirm that no proposal among them specifies a functional monetary control replacement for interest. The field has elaborated the diagnosis sufficiently. What it now requires is a technically coherent specification of what replaces interest as the monetary control instrument in a redesigned monetary architecture.

## 3 The functional architecture of interest: a decomposition

The six scholarly traditions mapped in Section 2 share a common limitation: they treat interest as a unified instrument whose problematic features — growth compulsion, wealth transfer, speculative bias — are inseparable from its necessary features — monetary supply control. If interest is a bundle of separable functions rather than an indivisible instrument, the reform problem changes fundamentally: instead of asking how to eliminate interest while preserving monetary stability, the question becomes which of interest's functions are necessary, which are eliminable, and what replaces the necessary ones. This section performs that decomposition.

*Table 1. The five functions interest bundles in the current monetary system, and their separability*

| Function | Description | Separable? |
|---|---|---|
| Monetary control instrument | Adjusts money supply by making credit expensive or cheap | NO - structurally necessary |
| Banking profit mechanism | Generates private income from the act of money creation | YES |
| Growth compulsion driver | Requires economy to expand to service compounding obligations | YES |
| Wealth transfer mechanism | Systematically transfers value from borrowers to lenders over time | YES |
| Speculative bias generator | Favours short-horizon financial returns over long-horizon investment | YES |

### 3.1 The five bundled functions of interest

The five functions listed in Table 1 are analytically separable despite being institutionally bundled in a single instrument. Monetary control is the one function that cannot be eliminated. Any functioning monetary system requires a mechanism to influence the quantity and velocity of money in circulation, and the reform question is therefore not whether to preserve this function but how.

The other four functions are all consequences of specific features of interest that are not intrinsic to monetary control. The banking profit mechanism is a consequence of interest accruing to the institution that creates the credit, and disappears in any architecture where money creation is a sovereign function and banks earn fee income for allocation and monitoring services. The growth compulsion driver, the wealth transfer mechanism, and the speculative bias generator all follow from a common feature of interest that has nothing to do with monetary control: the requirement of an additive return above principal, compounding over time. Growth compulsion follows because the economy must generate more money than was ever created to service the compounding obligation. Wealth transfer follows because guaranteed compound returns systematically shift value from borrowers to lenders and enable financial wealth to accumulate as power through passive ownership rather than productive contribution. Speculative bias follows because guaranteed compound returns on money systematically outcompete real productive investments whose returns are uncertain and whose horizons are long.

These three functions are therefore separable from monetary control not by argument but by construction. A monetary control instrument that does not carry additive returns above principal cannot generate any of them. The five-function bundle is the artifact of a particular design choice, not a structural necessity.

### 3.2 Why the bundling has been treated as necessary

Three intellectual traditions have contributed to the conflation of these separable functions. The first is the neoclassical time preference argument, that interest is the price balancing universal human preference for present over future goods. Behavioural economics consistently shows that temporal discounting is highly context-dependent (Thaler 1981; Loewenstein & Prelec 1992), and the historical record shows sophisticated commercial economies operating without compound interest for extended periods (Hartley & Kallis 2021). A more fundamental challenge is that positive time preference is itself partially endogenous to the interest-bearing system it justifies. Inflation heightens time preference (Gong 2006), and interest generates inflation substantially through the cost-push channel and monetary expansion. The time preference argument for interest may therefore be circular. This is offered as a hypothesis warranting further investigation rather than a conclusion.

The second is the capital allocation efficiency argument. Multiple institutional alternatives demonstrate the separability of this function from interest: the WIR Bank in Switzerland has operated an interest-free mutual credit system since 1934 with 62,000 members and approximately 3 billion Swiss francs in assets maintaining counter-cyclical stability through multiple financial crises (Stodder & Lietaer 2016); cooperative banks account for approximately one fifth of European Union bank deposits and loans; and the Islamic finance tradition has developed profit-sharing instruments, mudarabah and musharakah, demonstrating capital allocation without pure interest (El-Gamal 2006).

The third is the risk compensation argument, that interest incorporates a legitimate risk premium. This premium is separable from pure interest as a monetary control instrument. Risk differentiation can be performed through variable return rates calibrated to investment characteristics, productive cycle timescales, and social utility without requiring a guaranteed additive return on money as such. The Islamic finance tradition demonstrates this, as profit sharing contracts incorporate risk compensation through variable returns contingent on productive outcomes, without guaranteed compound interest.

### 3.3 Why the bundling is contingent, not natural

The historical and cross-cultural record reinforces the separability argument. The pre-enclosure European commons were collective governance systems that maintained fields, woodlands, and pastures across multiple generations through customary institutional rules rather than price signals (Ostrom 1990). This historical arrangement suggests that long-horizon ecological stewardship is achievable without the short-termism that interest-bearing finance generates. Fuders and Guerrero (2026) demonstrate through contemporary natural resource management cases that the same interest-driven dynamic operates in modern economies regardless of the production model. The bundling of interest's functions is therefore not a universal feature of monetary systems but a historically specific institutional arrangement that can be redesigned. The monetary control function alone remains the one function among the five that a monetary system genuinely requires and that therefore demands a replacement instrument rather than simple elimination. That instrument is the Principal Return Rate.

# 4 The Principal Return Rate system: mechanism, architecture and properties

The decomposition in Section 3 established that interest bundles five functions but only one of these, monetary control, is structurally necessary for a functioning monetary system. This section specifies the Principal Return Rate as the replacement instrument for that necessary function, proceeding through formal definition, the three-level nested architecture, the money supply mechanism, the inflation control mechanism, and a systematic comparison with existing alternatives.

## 4.1 Formal definition

The Principal Return Rate is defined as a percentage of the outstanding loan principal, set by the central bank as a national floor rate. This floor is adjusted in response to the aggregate inflation signal, upward when inflation exceeds target and downward when inflation falls below target or when economic conditions require support. Unlike the interest rate, which adds an obligation above the principal that the borrower must generate from productive activity, the PRR governs only the pace at which the original principal returns to its sovereign source. The borrower returns exactly what was allocated, at the rate the PRR determines, with no additive return above principal. This distinction is analytically fundamental. The interest rate is a price, which makes money more or less expensive to borrow, influencing demand for new credit and hence the quantity of money in circulation. The PRR is a velocity instrument: it governs how quickly existing money returns to the central bank's reserves, directly controlling the outstanding money stock without reference to the price of new credit. Three properties of this definition deserve immediate emphasis. First, no additive return above principal: money returns to the central bank at exactly the amount originally allocated, eliminating the wealth transfer mechanism. Second, no compounding obligation: the economy is not required to generate more money than was created, eliminating the growth compulsion. In the PRR system, capital cannot manifest as passive financial wealth accumulating power through compound interest returns. It can only exist as productive capital actively deployed in socio-ecologically purposeful activity, always in use and in circulation rather than held idle as a claim on future output. Compound interest has been the primary source of unpayable debts across every civilisation that allowed it to operate at scale (Hartley & Kallis 2021), and post-Keynesian theoretical work confirms its incompatibility with stationarity regardless of behavioural conditions (Cahen-Fourot

& Lavoie 2016). The problem that five millennia of regulatory intervention failed to eliminate through restriction, the PRR eliminates through design. Third, calibration to productive cycle timescales eliminates the speculative bias that interest generates by privileging short-horizon liquid assets over long-horizon productive investments.

## 4.2 The three-level nested architecture of the PRR system

The PRR system operates through a three-level nested architecture consistent with Ostrom's (1990) polycentric governance principle that effective governance requires institutional arrangements operating at the scale appropriate to the activity being governed. Level 1- the central bank sets a national minimum PRR floor, the primary macro monetary policy instrument: raising the floor accelerates money return to reserves, contracting the money supply; lowering the floor slows return, allowing expansion. The floor responds to the same inflation signal as the conventional interest rate but acts on contractually defined return schedules rather than behavioural responses to price signals, making transmission more direct and more predictable. Level 2- national techno-economic bodies, appointed through nationally accountable institutions, assess strategic national infrastructure allocations such as major energy systems, transport networks, national digital infrastructure. The assessment is done according to publicly codified criteria of productive utility and social-ecological value, with full transparency and independent audit infrastructure. This national level closes the political capture vulnerability at precisely the level where it is highest. Level 3- local techno-economic bodies appointed through locally accountable institutions operate on the same model as national bodies but at regional and local scale, setting PRRs not for individual loans but for predefined loan categories with published bands within which individual allocations fall, always at or above the national floor and transparent to borrowers before application.

The number and definition of categories is a matter of institutional design to be calibrated to local economic structure, not a fixed feature of the PRR mechanism itself. One illustrative scheme, presented here for concreteness, organises categories by productive cycle timescale and social-ecological utility. Category 1 covers long-horizon social-ecological investments such as renewable energy, ecological restoration, social housing, and public infrastructure, with PRR close to the national floor. Category 2 covers medium-horizon productive business investment, with PRR moderately above the floor. Category 3 covers short-horizon commercial activity, with PRR significantly above the floor. Category 4 covers consumer and household loans, with PRR set according to productive or social value. Category 5 covers speculative or purely financial activity, marked with the highest PRR, making such activity financially unattractive without explicitly prohibiting it. What is structurally required is only that local rates remain at or above the national floor and that bands are published and known to borrowers beforehand, exactly as in current commercial lending practice; the specific number and boundaries of categories may be adapted to the economic structure of the jurisdiction in which the system operates.

## 4.3 The money supply mechanism: demand-driven creation

Money creation in the PRR system is anchored to productive demand, exactly as in the current system. The unit of account is unchanged. The money supply mechanism operates through four stages. Stage 1: a borrower applies to a local BaaS institution, specifying investment purpose, which determines the loan category and applicable PRR band; the BaaS institution assesses creditworthiness while the local techno-economic body simultaneously assesses social-ecological utility; both assessments required. Stage 2: joint approval triggers a request from the BaaS institution to the central bank for sovereign money creation against the approved productive demand. Stage 3: the BaaS institution channels the sovereign money to the borrower under an agreed PRR schedule falling within the published band for the relevant category. Stage 4: principal returns to central bank reserves at the agreed PRR, ready for reallocation to new vetted demand, with the PRR floor adjusted in response to the aggregate inflation signal. This resolves the unit of account question open in sovereign money proposals since the Chicago Plan. Money is created against productive demand as it is now, but vetted for social-ecological utility, and returned through PRR velocity rather than interest cost.

## 4.4 Inflation control: the PRR as monetary policy instrument

Inflation operates through two structurally distinct channels in interest-bearing monetary systems, and the interest rate is implicated in generating both. When credit is cheap, expanded money supply generates demand-pull inflation: purchasing power rises faster than productive capacity, pulling prices upward. When credit is expensive, higher debt servicing costs generate cost-push inflation: prices rise not because of excess demand but because of increased production costs embedded in every debt-financed production process. The central bank raises rates to suppress demand-pull inflation; the rate rise generates cost-push inflation that constrains how far rates can be raised without triggering stagflationary recession. The instrument oscillates between generating one form of inflation and suppressing it through a mechanism that risks generating the other. It is a structural contradiction embedded in the use of a single price signal to manage two structurally distinct inflationary dynamics simultaneously.

The PRR resolves both channels simultaneously through a single architectural change. The demand-pull channel is eliminated because money supply is governed by PRR-determined return rates rather than by the price of new credit. The mechanism connecting cheap credit to money supply expansion does not exist in the PRR system. The cost-push channel is eliminated because the PRR imposes no cost burden on productive activity, there are no debt servicing costs to pass through to prices. Inflation is controlled through PRR floor adjustments by the central bank, operating on the same feedback signal as conventional interest rate policy but affecting inflation more directly and predictably. Unlike interest rate policy, which affects inflation indirectly through multiple behavioural and market channels whose responses are uncertain and variable, PRR floor adjustments affect inflation directly through contractually defined return schedules, making the policy effect more certain and predictable. The PRR system also avoids the zero lower bound trap that has constrained conventional monetary policy since the 2008 crisis. It is categorically different from zero interest rate policy (ZIRP). Zero or near zero central bank interest rates operated within an unchanged interest bearing private money creation architecture in which the growth compulsion remained fully intact. The PRR system replaces the architecture itself. What failed in ZIRP was a price signal adjustment within an unchanged system. What the PRR system proposes is the replacement of the system's monetary control instrument.

### 4.5 Comparison with existing alternatives

Table 2 systematically compares the PRR system against the four most significant existing proposals for monetary reform, using the six analytical dimensions established by decomposition in Section 3. The proposals are ordered by proximity to the PRR, from the closest to the most distant, to show where the PRR sits in relation to existing monetary reform proposals and what specific gap each of them leaves unaddressed.

*Table 2. PRR compared with existing monetary reform proposals*

| Proposal | Monetary control replaced? | Growth compulsion eliminated? | Wealth transfer eliminated? | Speculative bias eliminated? | Demand-driven money creation retained? | Direct inflation control? |
|---|---|---|---|---|---|---|
| Chicago Plan / Positive Money | NO; retains interest as residual control | NO; growth compulsion persists | NO; interest income persists | NO; speculative bias persists | UNSPECIFIED; sovereign money proposed but quantity anchor not defined | PARTIAL; relies on retained interest rate |
| Gesellian demurrage | NO; demurrage is a holding penalty, not a control instrument | NO; addresses hoarding incentive only; interest-bearing debt creation compulsion entirely unchanged | NO; interest on loans retained alongside demurrage | PARTIAL; reduces hoarding but not investment horizon bias | YES; existing money creation mechanism unchanged | NO; influences circulation behaviour but cannot directly target inflation |
| Islamic finance | NO; profit rate is additive return above principal | NO; operates within private bank money creation; monetary growth compulsion unchanged | NO; returns still flow from borrowers to capital holders | PARTIAL; profit-sharing reduces guaranteed return advantage but broader monetary speculative bias unchanged | YES; existing money creation mechanism unchanged | NO; relies on conventional monetary policy |
| Zero interest / ZIRP | NO; adjusts price signal within unchanged architecture | NO; growth compulsion persists regardless of rate level | NO: commercial lending rates remain positive; wealth transfer from borrowers to lenders persists | NO; speculative bias persists; asset price inflation amplified | YES; existing money creation mechanism unchanged | NO; exhausts conventional monetary policy space at the lower bound without providing a replacement mechanism |
| PRR (this article) | YES; return velocity replaces borrowing cost as control instrument | YES; no compounding obligation; money returns principal only | YES; no additive return above principal flows to capital holders | YES; PRR calibrated to productive cycle timescales | YES; demand-driven sovereign money creation explicitly retained by design | YES; central bank raises or lowers PRR floor in response to inflation signal; eliminates both demand-pull and cost-push channels |

The table reveals a precise pattern. Every existing proposal addresses some of the five bundled functions of interest, but no existing proposal replaces the monetary control function itself. The PRR system is the first proposal to replace the monetary control function directly and completely, closing the gap that the proposals compared here have identified but none has filled. It does so while retaining the one feature of the current system that works correctly: demand-driven money creation anchored to productive economic activity.

## 5 The institutional architecture of the PRR system

The PRR does not operate in isolation, it requires an institutional architecture capable of governing sovereign money creation, administering BaaS channelling, overseeing distributed techno-economic bodies, and maintaining the transparency infrastructure that makes the system publicly accountable. The purpose of this section is to demonstrate institutional operability, showing that each component has a credible governance design addressing the historical failure modes of public money creation.

### 5.1 Institutional Design for Sovereign Money Creation

Sovereign money creation in the PRR system is the exclusive mandate of the central bank, operating under a revised statutory framework that preserves monetary independence while extending independent oversight and public accountability. The central bank's existing mandate including price stability and financial system stability is retained and augmented with a third mandate that is social-ecological allocation coherence. PRR floor-setting decisions are made by the central bank's monetary policy committee through the same institutional process as current interest rate decisions such as periodic meetings, published minutes, and independent external scrutiny. The floor follows a rule-based adjustment: rising when aggregate inflation exceeds target and falling when inflation is below target or when economic conditions require support. The central bank's relationship to government is governed by the same operational independence principles that currently apply. What changes is not the governance structure of monetary policy but the instrument through which it is conducted.

### 5.2 The institutional transformation for Banking as a Service (BaaS)

BaaS institutions are the operational intermediaries of the PRR system meant to receiving sovereign money from the central bank, channelling it to approved borrowers, managing principal return schedules, and providing the full range of payment, account management, and financial advisory services. What they do not do is create money. Their revenue model shifts from interest-spread income to service fee income such as fees for credit assessment, payment processing, PRR schedule management, and investment matching. Existing commercial banks are the natural candidates for BaaS transformation since they possess the credit assessment expertise, customer relationships, and payment infrastructure that BaaS requires. The transformation redefines their function from money creators to money administrators without eliminating them.

The existing financial regulatory framework continues to apply in full, covering fair dealing with customers, consumer protection, and anti-money-laundering requirements. BaaS institutions maintain operational liquidity buffers sufficient to meet payment timing obligations. This is fundamentally simpler than conventional bank liquidity requirements because BaaS institutions do not fund long-term lending through short-term deposits. The maturity mismatch that makes bank runs structurally possible in the current system is therefore eliminated. Capital adequacy requirements in the Basel sense do not apply in the same form. Since BaaS institutions do not create money and do not bear credit risk on proprietary loan books, the credit risk on sovereign money allocations is borne by the central bank's sovereign money mandate.

## 5.3 Governance architecture of distributed polycentric techno-economic bodies

The distributed techno-economic bodies are the allocation governance layer tasked with assessing social-ecological utility of proposed investments, approving allocations within published PRR bands, and maintaining publicly codified criteria. Their governance design addresses four requirements simultaneously: civic legitimacy, technical competence, independence from political and financial capture, and local knowledge. Appointments occur through locally accountable institutions using open competitive processes with published competency criteria; members serve fixed terms with staggered renewal to prevent coordinated capture. Allocation criteria are codified at the national level through a publicly accountable legislative process and published in full. Local bodies apply nationally codified criteria to locally specific proposals but cannot modify them. This separation embodies Ostrom’s (1990) nested governance principle that is, higher levels set the rules, lower levels retain the discretionary judgment that local knowledge requires. Independence is maintained through three structural mechanisms: local rather than central ministerial appointment; full public recording on the digital transparency infrastructure; and an independent appeals mechanism with published decisions.

## 5.4 Digital transparency and public accountability

The digital transparency infrastructure makes every allocation decision, PRR schedule, and assessment outcome publicly visible and cryptographically immutable. It replaces the ex-post sample-based audit with continuous, full-scale, real-time public accountability. Every allocation is recorded on a distributed ledger accessible to any member of the public. The criteria against which each assessment was made are recorded alongside the outcome. This serves three accountability functions: visibility for citizens into how sovereign money is allocated; a real-time dataset for researchers and policymakers enabling continuous empirical evaluation; and full-population monitoring for regulators replacing sample-based retrospective audit. The transparency infrastructure records only assessments of investment purpose and productive utility and not social behaviour, political views, or personal characteristics of applicants. This distinction is architecturally enforced: the data fields recorded on the ledger are defined by the publicly codified allocation criteria, which contain no social or political variables.

### 5.5 The transition pathway

Three observations are sufficient here. First, the transition does not require sudden replacement. The PRR system can be introduced in parallel with the existing system, with sovereign money creation growing as a share of total credit while interest-bearing credit shrinks. Second, the global CBDC development currently underway provides the technical foundation for PRR implementation without requiring new infrastructure to be built from scratch. With 146 countries building sovereign digital money infrastructure, the architecture in its technical essentials is already a sovereign money infrastructure. Directing it toward PRR system governance rather than digital replication of the existing monetary order is an architectural decision, not a technical one. Third, the institutional components of the PRR system as proposed all have precedents in existing practice: central bank monetary policy committees, financial regulatory authorities, independent oversight bodies with publicly accountable appointment processes, and distributed ledger technology. What the PRR system requires is not the invention of new institutions but the coordination of existing institutional forms around a new monetary control instrument.

## 6 Addressing the known and anticipated critiques

The analytical decomposition in Section 3, the PRR mechanism in Section 4, and the institutional architecture in Section 5 collectively address most anticipated objections structurally. This section states the eight most significant objections in their strongest form and answers each directly.

### 6.1 The Hayekian knowledge problem

**Critique**: No central body can possess the information required for efficient credit allocation.

**Response**: The distributed polycentric architecture directly addresses the knowledge problem Hayek identified. The information is processed at the level where it exists, consistent with Ostrom's (1990) principle that effective governance requires institutional arrangements operating at the scale appropriate to the activity being governed. BaaS institutions assess borrower creditworthiness at the point of application, local techno-economic bodies assess the social-ecological utility of investments they know intimately, national bodies assess strategic national infrastructure, and the central bank adjusts the PRR floor to an aggregate inflation signal. The objection argues for the distributed watchdog design, not against it.

### 6.2 The political capture risk

**Critique**: Public bodies controlling credit allocation will be captured by political and financial interests.

**Response**: Four structural safeguards address this simultaneously; one, distribution across thousands of simultaneously operating local bodies makes systemic capture practically infeasible; two, digital transparency makes local capture immediately visible; three, nationally codified criteria constrain local discretion; and four, national bodies appointed through publicly accountable processes with published criteria and full audit trails prevent capture at the strategic infrastructure level. The current system, by contrast, is already in demonstrable financial sector capture with no equivalent safeguards

## 6.3 The inflation risk

**Critique**: Government control of money creation will produce rampant inflation.

**Response**: Section 4.4 demonstrates that the PRR resolves both inflation channels simultaneously, eliminating demand-pull by governing money supply through return velocity rather than credit pricing, and eliminating cost-push by imposing no cost burden on productive activity. The relevant historical precedents are instructive in the opposite direction; for example the 2008 global financial crisis and the 1991 Swedish banking crisis were not caused by government money creation but by the collapse of private interest-bearing money creation architectures that required perpetual growth to sustain themselves. This confirms that systemic catastrophe risk originates in the architecture the PRR replaces, not in sovereign money creation itself.

## 6.4 The ZIRP objection

**Critique**: Zero interest rates already failed so why would a PRR driven interest-free system succeed?

**Response**: ZIRP and PRR are categorically different. ZIRP adjusts a price signal to zero within an unchanged interest-bearing private money creation architecture. The growth compulsion remains fully intact, commercial banks continue creating money as interest-bearing debt, and the monetary architecture is preserved in all its essential features. The PRR replaces the architecture itself. What failed in ZIRP was a price signal adjustment within an unchanged system. What PRR proposes is the replacement of the system's monetary control instrument.

## 6.5 The unit of account and money quantity question

**Critique**: Without interest, what anchors the quantity of money created?

**Response**: Fully resolved by design. Money is created against vetted productive demand — exactly as in the current system. The unit of account is unchanged. The PRR governs the speed of return, not the logic of money creation. The quantity of money in circulation at any moment is determined by the aggregate of outstanding PRR-governed allocations, neither more nor less than the economy's vetted productive needs require.

## 6.6 The savings incentive problem

**Critique**: Without interest returns, nobody will save, removing lendable funds.

**Response**: The premise is empirically false. More than 97% of money is created by bank lending, not funded by household savings. Savings do not precede lending in the current system, lending precedes savings (Bank of England 2014). In the PRR system, money is created by the sovereign against productive demand, and the savings-as-funding chain does not exist. Active investors replace passive rentiers and returns derive from productive outcomes rather than guaranteed compound interest.

### 6.7 The international coordination problem

**Critique**: A single nation implementing PRR system faces arbitrage pressures from the global interest-bearing system.

**Response**: Acknowledged as the most structurally challenging transitional constraint. Three observations will suffice here: one, the PRR system is designed for gradual parallel introduction and not for sudden replacement; two, the CBDC infrastructure currently under development in 146 countries provides an unprecedented, coordinated entry point. Although the scale of simultaneous CBDC development provides a coordination opportunity that no previous monetary reform has offered, the detailed design of transition management requires empirical and political analysis which is beyond the scope of this article; and three, monetary architecture has changed through coordinated international agreement before, most recently at Bretton Woods.

### 6.8 The Chicago Plan comparison

**Critique**: The Chicago Plan already proposed sovereign money, so what does the PRR system add?

**Response**: The Chicago Plan and its descendants correctly identified private bank money creation as the structural problem but retained interest as the control instrument. Kumhof (2025) specifies a full architecture for the reformed system in which the sovereign administers four separate interest rates, on reserves, treasury credit, retail deposits, and wholesale deposits, together with a direct quantity rule for wholesale money. The control question is resolved, but through multiplying administered interest rates rather than replacing interest as the control instrument. The PRR fills this gap by specifying return velocity as the control instrument and PRR floor adjustment as the inflation response mechanism, completing what the Chicago Plan left unfinished.

## 7 Conclusions, limitations and future research

### 7.1 Main conclusions

The following conclusions emerge from this article's analysis and theoretical contributions.

Green growth through circular economy is structurally undermined by the interest-bearing monetary growth imperative however not by design failures at the production level. The service sector paradox confirms this empirically: CE-principled activities already generate approximately two thirds of GDP in advanced economies yet ecological overshoot persists, establishing the monetary architecture as the binding constraint regardless of how efficiently that value is generated. Interest bundles five functions of which only monetary control is structurally necessary. The bundling is historically contingent rather than logically necessary, as five millennia of cross-cultural evidence confirms. The PRR performs monetary control through return velocity without growth compulsion, wealth transfer and speculative bias. The PRR system retains demand-driven money creation while replacing only the instrument through which money supply is controlled. The PRR resolves both the demand-pull and cost-push inflation channels simultaneously, eliminating the structural contradiction embedded in interest rate policy as an inflation control instrument. The PRR closes the functional design gap in sovereign money theory unfilled since the Chicago Plan of 1933, confirmed as open by the most recent systematic reviews of the field. Finally, this article introduces the hypothesis that positive time preference, conventionally used to justify interest, is partially generated by the interest-bearing system itself. Inflation, which interest produces through cost-push and monetary expansion channels, heightens time preference. The argument that interest is necessary because people prefer present over future goods may therefore be circular.

## 7.2 Acknowledged limitations

The following two limitations are important to highlight here:

The first is the international coordination challenge: a single jurisdiction implementing the PRR system faces arbitrage pressures from the global interest-bearing system during any transition period, a constraint not fully resolved within the scope of a theoretical framework article. Coincidently, unprecedented scale of simultaneous CBDC development across 146 jurisdictions provides a coordination entry point that no previous moment of monetary reform has offered. However, the detailed design of managing this transition requires empirical and political analysis which is beyond this article's scope.

The second is transition pathway design: precise sequencing of institutional changes from the current monetary system to a PRR system depends on empirical and political conditions that vary by jurisdiction and cannot be fully specified before the institutional design work the PRR system requires.

## 7.3 Future research

The PRR system opens four new research directions.

First, formal stock-flow consistent modelling of the PRR system would provide the quantitative foundation this theoretical framework requires. Such modelling would demonstrate stability conditions, inflation control properties, and monetary control effectiveness under various economic conditions.

Second, detailed institutional design work on coordination architecture of BaaS-public watchdog, allocation criteria frameworks, and PRR determination framework would advance the implementation of the proposal.

Third, political economy analysis would address the most structurally challenging questions the proposal raises: the conditions under which transition becomes feasible, the mechanisms for international coordination, and the implications of the radical simplification of the financial system that the PRR enables.

Fourth, and most far-reaching, with interest removed from the monetary architecture, the research can now explore which of the 112 growth imperative mechanisms identified by Keysser et al. (2025) across 248 publications remain operative independently of the interest-bearing architecture and which are derivative of it. Testing the PRR system's claim to eliminate the monetary growth imperative and identifying remaining structural candidates for reform in a post-interest economy would be a substantive research programme.

Taken together, the PRR system proposal shows that the choice between monetary stability and ecological sustainability is not a genuine dilemma but an artefact of an interest-bearing monetary architecture whose growth compulsion has been naturalised rather than examined. The economy has already implicitly conducted the circular economy experiment at the scale of dominant value generation and the ecological outcomes confirm that no production model, however CE-principled, can overcome a monetary architecture that structurally compels aggregate growth. The primary intellectual obstacle to resolving this, the absence of a functional replacement for interest as the monetary control instrument, has been removed. Removal of the monetary growth imperative opens the possibility of an economy in which capital exists not as passive financial wealth accumulating power through interest but as productive capacity actively serving socio-ecological purposes. This is an economy that can grow in value and productive capacity without being structurally compelled to expand resource throughput and ecological impact. The practical work of building this alternative can now begin, alongside the theoretical work of mapping what remains to be resolved in a post-interest monetary architecture.